\documentstyle[12pt]{article}
\begin{document}
\setlength{\baselineskip}{23pt}
\newcommand{\vs}{\vspace{5pt}}
\newcommand{\vsa}{\vspace{10pt}}
\newcommand{\vsb}{\vspace{20pt}}
\newcommand{\lsim}{\begin{array}{c}<\vspace{-9pt}\\\sim\end{array}}
\newcommand{\gsim}{\begin{array}{c}>\vspace{-9pt}\\\sim\end{array}}
\newcommand{\invexp}{\rule{0pt}{10pt}}
\newcommand{\p}[1]{\frac{\partial}{\partial #1}}
\newcommand{\for}{\;\;\;\;{\rm for}\;\;\;\;}
\newcommand{\kld}{\left(1-\frac{N+8}{6}t\right)}
\newcommand{\lna}{\ln\left[1+z\kld^\frac{6}{N+8}\right]}
\newcommand{\lnb}{\ln\left[\frac{1}{3}+z\kld^\frac{6}{N+8}\right]}

\begin{titlepage}
\begin{flushright} UCLA/92/TEP/26 \\ July 1992 \\ hep-ph/9207252
\end{flushright}
\vsb
\begin{center}
{\large\bf Renormalization Group Improvement of the\\
Effective Potential in Massive O(N) Symmetric $\phi^4$ Theory}\vsa\vsb\\
Boris Kastening\footnote{email after 08/26/92: boris@ruunts.fys.ruu.nl}\vsb\\
{\it Department of Physics\\
University of California, Los Angeles\\
Los Angeles, California 90024}\vsb\vsb\\
{\bf Abstract}\\
\end{center}
The renormalization group is used to improve the effective potential
of massive ${\rm O}(N)$ symmetric $\phi^4$ theory. Explicit results are
given at the two-loop level.
\end{titlepage}

Recently techniques have been developed \cite{Ka} that allow to extend
renormalization group (RG) improvement of the effective potential (EP) in
one-component $\phi^4$ theory from the case of vanishing bare mass parameter
\cite{CoWe} to the massive case. In this paper one of those methods is
extended to the ${\rm O}(N)$ symmetric model. This is interesting because
the model contains Goldstone bosons if ${\rm O}(N)$ is spontaneously broken.

Also two different types of logarithms appear in the course of infinite
renormalization since the Higgs and Goldstone boson masses depend
differently on the classical background field. In more realistic theories
typically several such logarithms appear, e.g.\ through the additional
presence of fermions and gauge bosons that get their masses through Yukawa
and gauge couplings, respectively, when the Higgs field acquires a nonzero
vacuum expectation value. Since RG improvement at the one-, two-,
\ldots~loop level amounts to summing up leading, next-to-leading,
\ldots~logarithmic terms of the potential, one has to worry about
the meaning of that statement when more than one kind of logarithm is
present.

Our model is defined by the Lagrangian
\begin{equation}
{\cal L}=\frac{1}{2}\partial_\mu\phi_i\partial^\mu\phi_i-V_0,\,\,\,\,\,\,\,\,
V_0=\frac{\lambda}{4!}\phi^4+\frac{m^2}{2}\phi^2,
\end{equation}
where $\phi^2\equiv\phi_i\phi_i$ and $i=1,\ldots,N$. The Feynman rules
are easily worked out and one can compute the one-, two-, \ldots~loop
contribution to the EP, e.g.\ making use of vacuum graphs in a shifted
theory \cite{Ja}.

In order to RG improve the EP, we first need the unimproved potential.
With dimensional regularization \cite{tHVeBoGiAs} and the
$\overline{\rm MS}$-scheme \cite{tHBaBuDuMu}, a scheme used throughout
this paper, the one-loop contribution $V_1$ to the EP is
easily seen to be given by
\begin{equation}
\label{onv1}
(4\pi)^2V_1=\frac{m_H^4}{4}\left(\ln\frac{m_H^2}{\mu^2}-\frac{3}{2}\right)
+(N-1)\frac{m_G^4}{4}\left(\ln\frac{m_G^2}{\mu^2}-\frac{3}{2}\right),
\end{equation}
where $m_H^2=\frac{\lambda}{2}\phi^2+m^2$, $m_G^2=\frac{\lambda}{6}\phi^2+m^2$,
and $\mu$ is the renormalization scale.

The two-loop contribution is considerably harder to determine. However,
recently this has been achieved \cite{FoJo} and the result is\footnote{As one
can easily convince oneself, there is a typo in the formula given in
\cite{FoJo}. I am grateful to Tim Jones for pointing this out to me.}
\begin{eqnarray}
\label{onv2}
(4\pi)^4V_2\;=&&\frac{1}{8}\lambda^2\phi^2m_H^2\left(\ln^2\frac{m_H^2}{\mu^2}
-4\ln\frac{m_H^2}{\mu^2}+8\Omega(1)+5\right)
+\frac{1}{8}\lambda m_H^4\left(\ln\frac{m_H^2}{\mu^2}-1\right)^2\nonumber\\
&+&(N-1)\left\{\frac{1}{72}\lambda^2\phi^2\left[(m_H^2+2m_G^2)
\left(\ln^2\frac{m_G^2}{\mu^2}-4\ln\frac{m_G^2}{\mu^2}
+8\Omega\left(\frac{m_H^2}{m_G^2}\right)+5\right)\right.\right.\nonumber\\
&&\left.\hspace{103pt}+2m_H^2\ln\frac{m_H^2}{\mu^2}
\left(\ln\frac{m_G^2}{\mu^2}-2\right)\right]\nonumber\\
&&\left.\hspace{59pt}+\frac{1}{12}\lambda m_H^2m_G^2
\left[\ln\frac{m_H^2}{\mu^2}\ln\frac{m_G^2}{\mu^2}-\ln\frac{m_H^2}{\mu^2}
-\ln\frac{m_G^2}{\mu^2}+1\right]\right\}\nonumber\\
&+&(N^2-1)\frac{1}{24}m_G^4\left(\ln\frac{m_G^2}{\mu^2}-1\right)^2,
\end{eqnarray}
where $\Omega$ is defined by
\begin{equation}
\Omega(x)\equiv\left\{\begin{array}{ll}
\displaystyle\frac{\sqrt{x(4-x)}}{x+2}\int_0^{\arcsin(\frac{1}{2}\sqrt{x})}
\ln(2\sin t)dt&\for x\leq 4\vsa\\
\displaystyle\frac{\sqrt{x(x-4)}}{x+2}\int_0^{{\rm
arcosh}(\frac{1}{2}\sqrt{x})}
\ln(2\cosh t)dt&\for x>4
\end{array}\right..
\end{equation}
\begin{sloppypar}
Since (see \cite{FoJo})
$\lim_{x\rightarrow\infty}\{\Omega(x)
-[\frac{1}{8}\ln^2x+\frac{1}{4}\zeta(2)]\}=0$,
$V_2$ is finite at the tree-level minimum (where $m_G^2=0$), as is $V_1$.
Let us define $\widetilde{V}\equiv V/\phi^4$, $y\equiv\ln(m_H^2/\mu^2)$,
and $z\equiv 2m^2/(\lambda\phi^2)$. Because $\mu$ appears in the $n$-loop
contribution to the EP only in terms proportional to
$\ln^{k_H}(m_H^2/\mu^2)\ln^{k_G}(m_G^2/\mu^2)$, where $0\leq k_H+k_G\leq n$,
the EP as computed loop by loop can be written as
\end{sloppypar}
\begin{equation}
\label{onlpbylp}
\widetilde{V}=\lambda\left(\frac{1}{24}+\frac{z}{4}\right)
+\sum_{L=1}^\infty\frac{\lambda^{L+1}}{(4\pi)^{2L}}\sum_{n=0}^Ly^ng_{Ln}(z),
\end{equation}
where we have made use of the fact that every loop introduces another factor
of $\lambda$, when writing $\widetilde{V}$ in terms of $\lambda$, $y$,
and $z$.

After rewriting the renormalization group equation (RGE)
\begin{equation}
\label{onrge}
\left(\mu\p{\mu}+\beta(\lambda)\p{\lambda}+\gamma_m(\lambda)m^2\p{m^2}
-\gamma(\lambda)\phi\p{\phi}\right)V(\lambda,m^2,\phi,\mu)=0
\end{equation}
into an equation for $\widetilde{V}$ in terms of $\lambda$, $y$, and $z$,
it is straightforward to show that (\ref{onlpbylp}) with the first few
$g_{Ln}$ determined by (\ref{onv1}) and (\ref{onv2}) fails to obey this RGE.
As in the $N=1$ case \cite{Ka} the problem can be cured by introducing
a suitable $\mu$-independent tree-level constant into the potential.
Then (\ref{onlpbylp}) becomes
\begin{eqnarray}
\label{onvfull}
\widetilde{V}_{\rm full}&=&\lambda\left(\frac{1}{24}+\frac{z}{4}\right)
+\sum_{L=1}^\infty\frac{\lambda^{L+1}}{(4\pi)^{2L}}\sum_{n=0}^Ly^ng_{Ln}(z)
+z^2\sum_{k=1}^\infty b_k\lambda^k\nonumber\\
&\equiv&\sum_{L=0}^\infty\frac{\lambda^{L+1}}{(4\pi)^{2L}}
\sum_{n=0}^Ly^ng_{Ln}(z)+z^2\sum_{k=1}^\infty b_k\lambda^k\nonumber\\
&\equiv&\sum_{k=1}^\infty\lambda^kf_k(z,t),
\end{eqnarray}
where $t\equiv\lambda y/(4\pi)^2$, the $b_k$ are to be determined
by demanding consistency with the unimproved $k$-loop potential and
\begin{equation}
\label{onfk}
f_k(z,t)\equiv(4\pi)^{2(1-k)}\sum_{n=0}^\infty t^ng_{k+n-1,n}(z)+b_kz^2.
\end{equation}
The tree-level potential, represented by $g_{00}$ in (\ref{onvfull}),
is part of $f_1$.

If we write $\ln(m_G^2/\mu^2)$ as $\ln(m_G^2/m_H^2)+\ln(m_H^2/\mu^2)$,
then $f_k$ contains all $k$-th leading powers in $\ln(m_H^2/\mu^2)$, i.e.\
for every $n$ it contains the terms proportional to
$\ln^{n-k+1}(m_H^2/\mu^2)$ of the $n$-loop contribution to the EP.

The RGE (\ref{onrge}) can be rewritten \cite{Ka} as recursive differential
equations for the functions $f_k$:
\[
(4\pi)^2\sum_{k=1}^{L}\left\{\beta_{k+1}t\p{t}
-(\beta_{k+1}-\alpha_k-2\gamma_k)z\p{z}+[(L-k+1)\beta_{k+1}
-4\gamma_k]\right\}f_{L-k+1}\nonumber
\]
\begin{equation}
\label{onrecursive}
-2\frac{\partial f_L}{\partial t}
+\sum_{k=1}^{L-1}\frac{\beta_{k+1}-2\gamma_k+z\alpha_k}{1+z}
\frac{\partial f_{L-k}}{\partial t}=0,
\end{equation}
where the $\alpha_k$, $\beta_k$, and $\gamma_k$ are defined by
\begin{equation}
\gamma_m\equiv\sum_{k=1}^{\infty}\alpha_k\lambda^k,\;\;\;\;\;\;\;\;
\beta\equiv\sum_{k=2}^{\infty}\beta_k\lambda^k,\;\;\;\;\;\;\;\;
\gamma\equiv\sum_{k=1}^{\infty}\gamma_k\lambda^k.
\end{equation}
The boundary conditions for the equations (\ref{onrecursive}) are given by
(\ref{onfk}) at $t=0$, i.e.\
\begin{equation}
\label{onbc}
f_k(z,0)=(4\pi)^{2(1-k)}g_{k-1,0}(z)+b_kz^2.
\end{equation}
To fix $b_k$ we demand $f_k$ to be consistent with $g_{k1}$ which in turn
can be extracted from the $k$-loop contribution to the EP, $V_k$. With
(see e.g.\ \cite{FoJo,5loops})
\begin{equation}\begin{array}{lclcll}
\alpha_1=\!\!\!\!&\displaystyle\frac{N+2}{3(4\pi)^2},\;\;\;\;\;\;
&\beta_2=\!\!\!\!&\displaystyle\frac{N+8}{3(4\pi)^2},\;\;\;\;\;\;
&\gamma_1=\!\!\!\!&\displaystyle 0,\vsa\\
\alpha_2=\!\!\!\!&\displaystyle-\frac{5(N+2)}{18(4\pi)^4},\;\;\;\;\;\;
&\beta_3=\!\!\!\!&\displaystyle-\frac{3N+14}{3(4\pi)^4},\;\;\;\;\;\;
&\gamma_2=&\!\!\!\!\displaystyle\frac{N+2}{36(4\pi)^4},
\end{array}\end{equation}
we are ready now to compute $f_1$ and $f_2$.

With
\begin{equation}
g_{00}(z)=\lambda\left(\frac{1}{24}+\frac{z}{4}\right)
\end{equation}
in the boundary condition (\ref{onbc}) we can solve (\ref{onrecursive})
for the case $L=1$. Upon expanding the resulting expression in $t$ and
matching the linear term with
\begin{equation}
g_{11}(z)=\frac{1}{16}(1+z)^2+\frac{1}{16}(N-1)(1/3+z)^2
\end{equation}
gotten from (\ref{onv1}), we get $b_1=3N/[8(N-4)]$ and
\begin{equation}
\label{onf1}
f_1(z,t)=\frac{1}{24}\kld^{-1}+\frac{z}{4}\kld^{-\frac{N+2}{N+8}}
+\frac{3Nz^2}{8(N-4)}\kld^{-\frac{N-4}{N+8}}.
\end{equation}
Note that the $\phi$-dependent part of $f_1\phi^4$ remains finite for $N=4$.
As a further check of (\ref{onf1}) one can determine the $t^2$-term in $f_1$
and compare it with $g_{22}$ gotten from (\ref{onv2}) and find agreement.

Next we compute $f_2$. We can extract
\begin{equation}
g_{10}(z)=-\frac{3}{32}(1+z)^2+\frac{1}{16}(N-1)(1/3+z)^2
\left[\ln\left(\frac{1/3+z}{1+z}\right)-\frac{3}{2}\right]
\end{equation}
from (\ref{onv1}) and use it in the boundary condition (\ref{onbc}) to solve
(\ref{onrecursive}) for $L=2$. Upon expanding the resulting expression in
$t$ and matching the linear term with
\begin{eqnarray}
g_{21}(z)\;=&&(N^2-1)\frac{(1/3+z)^2}{48}
\left[\ln\left(\frac{1/3+z}{1+z}\right)-1\right]\nonumber\\
&+&(N-1)\left[\frac{(1/3+z)(7/3+z)}{48}\ln\left(\frac{1/3+z}{1+z}\right)
-\left(\frac{z^2}{24}+\frac{5z}{36}+\frac{13}{216}\right)\right]\nonumber\\
&-&\frac{(1+z)(5+z)}{16},
\end{eqnarray}
which can be extracted from (\ref{onv2}), we get
$b_2=-N(N+8)/[4(4\pi)^2(N-4)(N+2)]$ and

\begin{equation}
\label{onf2}
(4\pi)^2f_2(z,t)\;=\!\!
\begin{array}[t]{rl}
&\left\{-\frac{N+8}{96}\right.-\frac{N+2}{432}t\vs\\
&\;\;-\frac{N^2-2N-20}{144(N+8)}\ln\kld+\frac{1}{16}\lna\vs\\
&\;\;+\frac{N-1}{144}\lnb-\left.\frac{N+8}{144}\ln(1+z)\right\}
\kld^{-2}\vsb\\
\!\!+\!\!\!\!&\left\{-\frac{N+2}{16}\right.+\frac{(N+2)(N+3)}{12(N+8)}t\vs\\
&\;\;-\frac{(N+2)(N^2-2N-20)}{24(N+8)^2}\ln\kld+\frac{1}{8}\lna\vs\\
&\;\;+\frac{N-1}{24}\lnb-\left.\frac{N+2}{24}\ln(1+z)\right\}
z\kld^{-2\frac{N+5}{N+8}}\vsb\\
\!\!+\!\!\!\!&\left\{-\frac{N(3N^2+2N+40)}{32(N+2)(N-4)}\right.
+\frac{N(N+2)(13N+44)}{48(N+8)(N-4)}t\vs\\
&\;\;-\frac{N(N^2-2N-20)}{16(N+8)^2}\ln\kld+\frac{1}{16}\lna\vs\\
&\;\;+\frac{N-1}{16}\lnb-\left.\frac{N}{16}\ln(1+z)\right\}
z^2\kld^{-2\frac{N+2}{N+8}}.
\end{array}
\end{equation}
Again one can show that the $\phi$-dependent part of $f_2\phi^4$ remains
finite for $N=4$.

Now instead of choosing $y\equiv\ln(m_H^2/\mu^2)$ as the relevant logarithm,
one could have taken $x\equiv\ln(m_G^2/\mu^2)$ as another natural choice.
With $s\equiv\lambda x/(4\pi)^2$, it is easy to see that then our result
would have been
\begin{equation}
\widetilde{V}_{\rm full}=\sum_{k=1}^\infty\lambda^k\widetilde{f}_k(z,s)
\end{equation}
with the modified functions
\begin{eqnarray}
\widetilde{f}_1(z,s)&=&f_1(z,s),\\
(4\pi)^2\widetilde{f}_2(z,s)&=&(4\pi)^2f_2(z,s)-\ln\left(\frac{1/3+z}{1+z}
\right)\p{s}f_1(z,s),
\end{eqnarray}
and so on. For large fields sufficiently short of the Landau pole at
$t=6/(N+8)$, $x\approx y$ and $s\approx t$ hold and our RG improved
approximations to the potential do not change much if we use $s$ instead of
$t$. However, if there is spontaneous symmetry breaking due to negative
$m^2$, the potential changes completely around the tree-level minimum.
In fact, the second derivative of the unimproved one-loop potential diverges
there as does the second derivative of the one-loop improved result, if
we use $\lambda$ and $s$. If we expand in powers of $\lambda$ and $t$, this
is true starting at the two-loop level. This indicates that we should not trust
our result for fields around or smaller than that minimum. Neither should
one trust the unimproved result (\ref{onv2}) there. The reason is, of course,
the presence of infrared divergences due to Goldstone bosons.

In summary, we have used the renormalization group to obtain an improved
version of the effective potential in $O(N)$ symmetric $\phi^4$ theory.
Eqs.~(\ref{onf1}) and (\ref{onf2}) represent our results at the one- and
two-loop level. The benefit of the improvement is for large fields
sufficiently short of the Landau pole, while for fields around or smaller
than the tree-level minimum infrared divergences make both the unimproved
and the improved result untrustworthy.

\section*{Acknowledgements}
I am grateful to Roberto Peccei for useful comments on the manuscript
and to Tim Jones for helpful communication. This work was supported
in part by the Department of Energy under Contract No.~DE-AT03-88ER 40383
Mod A006-Task C.

\end{document}